\documentclass[%
 reprint,
superscriptaddress,
 amsmath,amssymb,
 aps,
 prl]{revtex4-2}

\usepackage{graphicx}% Include figure files
\usepackage{dcolumn}% Align table columns on decimal point
\usepackage{bm}% bold math
\usepackage{xcolor}
\usepackage{ulem}
\newcommand{\corr}[1]{\begin{color}{orange}#1\end{color}}
\newcommand{\corra}[1]{\begin{color}{red}#1\end{color}}
\newcommand{\tsl}[1]{\begin{color}{orange}#1\end{color}}
\newcommand{\tslc}[1]{\begin{color}{magenta}#1\end{color}}
\renewcommand{\sout}[1]{\null}
\renewcommand{\corr}[1]{#1}
\renewcommand{\tsl}[1]{#1}
\renewcommand{\corra}[1]{\null}
\renewcommand{\tslc}[1]{\null}

\begin{document}

\title{Experimental feasibility of molecular two-photon absorption with isolated time-frequency-entangled photon pairs}

\author{Tiemo Landes}
\email{oqt@uoregon.edu}
\affiliation{Department of Physics and Oregon Center for Optical, Molecular, and Quantum Science, University of Oregon, Eugene, Oregon 97403, USA}
\affiliation{Department of Chemistry and Biochemistry, Center for Optical, Molecular and Quantum Science, University of Oregon, Eugene, Oregon 97403, USA}
\author{Markus Allgaier}
\affiliation{Department of Physics and Oregon Center for Optical, Molecular, and Quantum Science, University of Oregon, Eugene, Oregon 97403, USA}
\author{Sofiane Merkouche}
\affiliation{Department of Physics and Oregon Center for Optical, Molecular, and Quantum Science, University of Oregon, Eugene, Oregon 97403, USA}
\author{Brian J. Smith}
\affiliation{Department of Physics and Oregon Center for Optical, Molecular, and Quantum Science, University of Oregon, Eugene, Oregon 97403, USA}
\author{Andrew H. Marcus}
\affiliation{Department of Chemistry and Biochemistry, Center for Optical, Molecular and Quantum Science, University of Oregon, Eugene, Oregon 97403, USA}
\author{Michael G. Raymer}
\affiliation{Department of Physics and Oregon Center for Optical, Molecular, and Quantum Science, University of Oregon, Eugene, Oregon 97403, USA}

\date{\today}% It is always \today, today,
             %  but any date may be explicitly specified

\begin{abstract}
Entangled photon pairs have been promised to deliver a substantial quantum advantage for two-photon absorption spectroscopy. However, recent work has challenged the previously reported magnitude of quantum enhancement in two-photon absorption. Here, we present an experimental comparison of sum-frequency generation and molecular absorption, each driven by isolated photon pairs. We establish an upper bound on the enhancement for entangled-two-photon absorption in Rhodamine-6G, which lies well below previously reported values.
\end{abstract}

%\keywords{Suggested keywords}%Use showkeys class option if keyword
                              %display desired
\maketitle

\section*{Introduction}

Quantum states of light have been employed as a resource across disciplines, from telecommunications to metrology to spectroscopy, with the goal of enhancing existing techniques with a \textit{quantum advantage}.
In the past decade, the use of time-frequency entangled photon pairs (EPP) for two-photon absorption (TPA) has seen growing interest in this context \cite{Schlawin2016,Szoke2020}. Such photon entanglement has been promised to enhance TPA for molecular and atomic spectroscopy in an attempt to circumvent the strong optical powers typically required to observe TPA. Several TPA experiments aimed at observing a quantum advantage using EPP have been performed using a variety of molecular systems \cite{Goodson2006,Valencia2017,Mikhaylov2020,Goodson2017,Goodson2018,Valencia2017,Tabakaev2020}. Recent publications have reported direct observation of entangled TPA-induced fluorescence \cite{Goodson2017,Goodson2018,Tabakaev2020}, although with limited yield. The reported strength of TPA with EPP is in many cases comparable to single-photon absorption within a few orders of magnitude, which translates to a quantum enhancement of more than 10 orders of magnitude when compared to TPA with nonentangled light. This enhancement has recently been challenged by new experimental results, suggesting that any enhancement is too small to be observed in \corr{realistic} measurements, where expected signals are extremely small to begin with \cite{parzuchowski2020}.

Theoretically, many aspects of \corr{entangled two-photon absorption} (ETPA) have been described, namely the interaction time scales arising from time-frequency entanglement \cite{Fei1997,Dayan2007}. A comprehensive tutorial of the topic can be found in \cite{Schlawin2017}. \corr{Most theoretical work on ETPA had neglected to account for the rate at which signal can be obtained in different regimes} with regards to flux, optical pulse duration, and the amount of time-frequency entanglement. Recent calculations cast doubt over previously published experimental results and thus prompt further investigation into the magnitude of \corr{TPA} enhancement through the use of EPP \cite{raymer2020}. That treatment emphasizes the separate roles played by spectral correlations and by photon-number correlations. These new predictions suggest that large enhancement of the TPA cross section through time-frequency entanglement \corr{occurs only} for narrowband transitions \corr{which are not typically present in molecular solutions}. The advantage provided by \corr{photon-}number correlations is operable only in the limit of extremely low photon flux where signals are practically unobservable below common detection thresholds. \corr{It has been predicted that the quantum enhancement factor (\(QEF\)) for broadband transitions only depends on EPP bandwdith \(B\) and photon flux \(F\), \(QEF \approx B/F\) \cite{raymer2020}. While the enhancement can be several orders of magnitude, it is predicted to be insufficient for observing TPA-induced fluorescence at typical EPP flux.} These novel predictions are consistent with results of a recent survey of ETPA in a variety of dyes \cite{parzuchowski2020}, where a notable lack of any TPA-induced fluorescence signal was taken as indicating little quantum enhancement of TPA. That survey makes several convincing arguments regarding the nature of the linear photon-flux scaling behavior that is often associated with an entangled-two-photon process. However, additional experimental steps are warranted in the absence of a fluorescence signal. Adverse effects such as optical dispersion, spatial overlap of photon pairs in their focal volume, and the precise photon-flux scaling behavior associated with two-photon processes and two-photon beams need to be considered.

\corr{In this letter,} we present experimental results comparing sum-frequency generation \corr{(SFG) and molecular TPA in Rhodamine 6G using the same setup}, each driven by isolated photon pairs. The results, along with comparison with the recent theory \cite{raymer2020},  provide a convincing limit on quantum advantage in molecular TPA. \sout{Using a single-mode fiber as a test volume we measure a lower bound on the total EPP flux. This allows us to establish that} For our experiment the enhancement of TPA \tsl{in Rhodamine 6G} by EPP is no more than \(1.5(\pm0.3)\times10^{5}\)-times greater than what is predicted by our theory, several orders of magnitude below previously reported values.

A review of previously employed experimental methods serves as a guide for designing benchmark experiments able to verify the results in \cite{raymer2020}.
Valuable test-bed systems for the study of the general nature of two-photon interaction have been TPA in atoms \cite{Georgiades1995,Dayan2004} as well as SFG in second-order nonlinear crystals \cite{Dayan2005,Odonnell2020}. The role of time-frequency entanglement in SFG can be understood in the framework of time-frequency modes \corr{\cite{fabre2020}}. TPA experiments on atomic systems revealed the importance of photon number correlations. SFG serves well to study the photon-flux scaling behavior \corr{of TPA} when EPP are employed. \corr{In molecular TPA, observing }scaling behavior of the absorption and fluorescence that is quadratic in power of an incident \corr{coherent-state} beam is an essential \corr{piece of evidence} for ruling out single-photon processes \sout{, and these experiments show that a log-log plot best serves to evaluate this scaling. This practice has also been adopted by \cite{parzuchowski2020}.} \corr{We note} that none of the observations of ETPA in molecular samples mentioned here \cite{Goodson2006,Valencia2017,Mikhaylov2020,Goodson2017,Goodson2018,Tabakaev2020} underwent this rigorous scaling test. In all \corr{previous} experiments considered here, the EPP are generated by spontaneous parametric down-conversion (SPDC) driven by a pump laser with frequency equal to the TPA transition under study. \sout{Therefore, in molecules with non-zero single-photon absorption cross section at any wavelength present in the experiment (such as Rhodamine B \cite{Valencia2017}), such a test is indispensable to rule out processes such as single-photon absorption. Even small amounts of scattered \sout{photons, or residual SPDC pump} light can easily obscure or mimic weak signals apparently originating from a two-photon process.}\corra{(Explained elsewhere)} \corr{In the low-flux regime, where theory predicts illumination by EPP should produce a higher TPA signal compared with classical laser light, the crucial distinction lies in the scaling of the TPA signal arising from EPP. The TPA signal should scale linearly with the EPP pump power, and quadratically when the EPP is directly attenuated, as shown for SFG \cite{Dayan2005}.} A second \corr{condition necessary to provide convincing evidence of enhanced TPA} is \sout{the presence of} a large amount of time-frequency entanglement in the EPP state\sout{ to provide convincing evidence of an enhancement of TPA, which has to be considered for designing an appropriate source of EPP.}. \corr{Pumping SPDC with a spectrally narrow laser} results in a high degree of spectral entanglement and temporal isolation of pairs \sout{is required to avoid overlapping of uncorrelated pairs.} Therefore, a continuous-wave (CW) pumped SPDC source, as used in \cite{Tabakaev2020}, is preferable over a \corr{pulse-pumped SPDC} source for observing quantum enhancement of TPA, \corr{especially for generating large numbers of isolated pairs.} \sout{To put it simply, one definitely requires spectral entanglement to demonstrate an effect based thereon. Pulsed EPP sources with limited degree of entanglement such as the type-II SPDC employed by Goodson et al. may not be sufficient. An \sout{second} advantage of CW-pumped sources is that they allow generation of a \corr{high rate} of isolated EPP.} \sout{While we achieve average photon \corr{production rates} of the order of \(10^9s^{-1}\) in the isolated-pair regime, a pulsed source like the one used by Parzuchowski et al. \cite{parzuchowski2020} would produces roughly \(10^2\) overlapping photons per pulse at the same time-averaged flux.} Lastly, \corr{careful} characterization of the interaction volume and EPP flux in that volume are essential for \corr{precisely bounding the quantum enhancement of TPA with EPP.}\\

\corr{In what follows,} we first describe the experimental setup \corr{and} a study of TPA scaling behavior \corr{of SFG} with flux for both EPP and classical light. After detector and flux calibration procedures are explained, we \corr{determine} the expected TPA fluorescence signal from the molecular sample. Comparison with detection threshold allows us to establish an upper bound on the enhancement of EPP.

\section*{Experimental setup}

\begin{figure}
    \centering
    \includegraphics[width=0.4\textwidth]{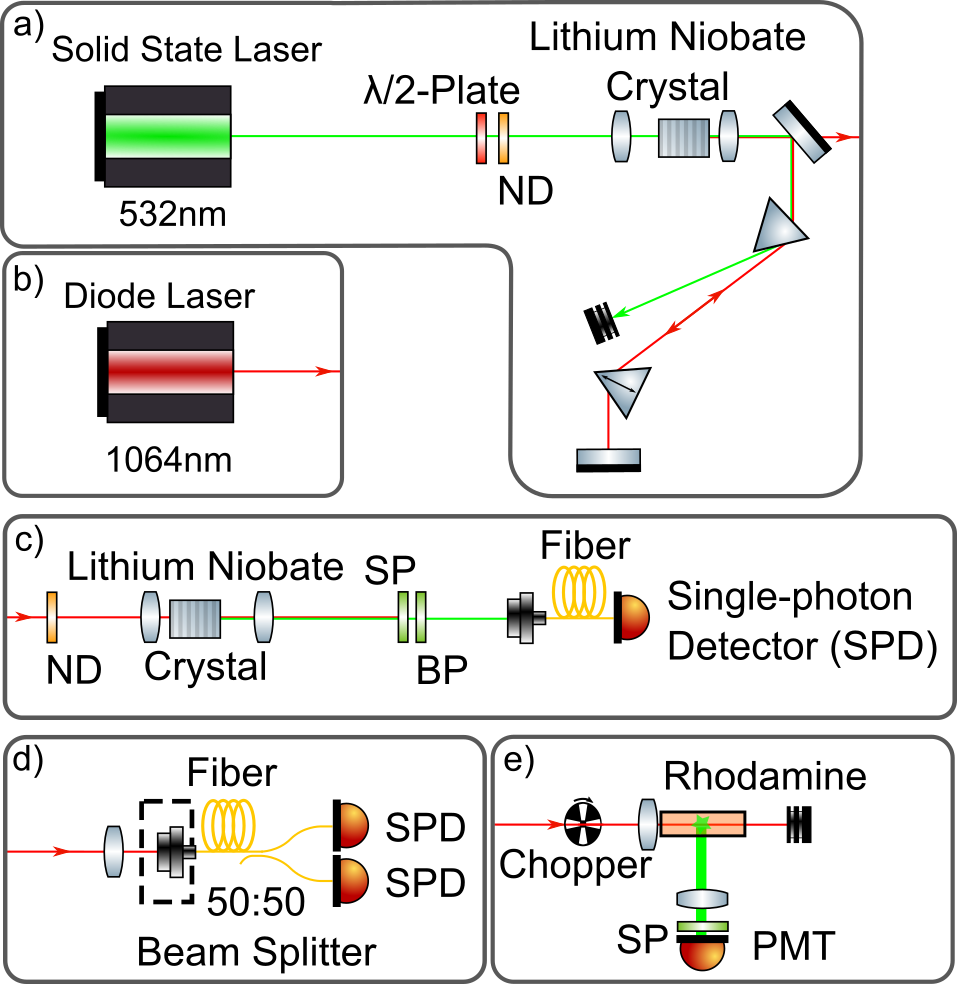}
    \caption{For measurement of scaling behavior, either the entangled photon pairs from a Lithium Niobate source (a), or coherent light from a diode laser (b) are coupled through another Lithium Niobate crystal (c). The SFG is then coupled into a single-mode fiber, which is connected to a single photon detector. ND (Neutral density) filters in front and behind the EPP source are used to compare scaling behavior. Calibration of the fluorescence collection apparatus is performed by observing the two-photon absorption (e) from coherent light (b). We then replace the molecular sample with a single mode fiber in the same place (d) to obtain a lower bound on the number of photon pairs emitted by (a). PMT: Large-area photo multiplier tube, SP: Shortpass, BP: Bandpass}.
    \label{fig:setup}
\end{figure}

The experimental setup sketched in Figure \ref{fig:setup} has two parts: A light generation (panels a-b) and a detection block (c-e). Two different light sources are available. A type-0 SPDC source, pumped by a 532\,nm-CW laser provides EPP centered around 1064\,nm (panel a). A prism pulse compressor is employed to compensate for second-order dispersion. Details on EPP source and pulse compressor are explained in the supplementary material. Dispersion compensation is later verified by maximizing the signal from the subsequent SFG (panel c),\sout{, thus ensuring that our experiment does not suffer from the adverse effects of dispersion.} \corr{which also removes the 532\,nm pump light}. Residual scattered 532\,nm light is removed with several longpass filters. As a calibration source and for comparison of scaling behavior we employ a CW diode laser at 1064\,nm (panel b).

In the first experiment (panel c) we study SFG by isolated EPP \corr{to confirm that all requirements for successful ETPA are met}. We focus \corr{collimated} EPP \corr{emerging from the prism compressor} into a second identical nonlinear crystal \corr{with a second achromatic lens with the same focal length of 100\,mm}. After the SFG crystal, \corr{residual EPP} are removed using a pair of short-pass and a band-pass-filters. Light generated via SFG is coupled into a single-mode fiber and detected on an avalanche photo diode\sout{(\textit{Laser Components Count Blue})}. This experiment is repeated with the 1064\,nm CW diode laser \corr{replacing the EPP}. We attenuate the diode laser, the SPDC and the SPDC pump in order to verify scaling behavior of the SFG output counts with \corr{input} IR power\sout{ for both SPDC photons and CW light}.

\corr{To examine the behavior of molecular TPA with} \tsl{our} \corr{EPP source, we first bound the EPP flux passing through the molecular sample. This is done by placing a single mode fiber at the focus of the sample cell.} The fiber assembly is designed to be removed without misaligning the focusing lens, or any prior optics. The fiber-coupled EPP are split probabilistically using a 50:50 fiber beamsplitter and detected on superconducting nanowire single photon detectors \sout{(\textit{iD Quantique})}. The absolute coupling efficiency is far from unity, and only a \corr{fraction of EPP is collected}. The measured coincidence rate serves to establish a lower bound on the number of pairs in the interaction volume within the cuvette. This provides a much tighter bound on the delivered pair flux \corr{than simply measuring power or single-photon counts in free space,} especially in the presence of loss \(L\), \sout{, pair rates ought not be estimated from power,} where \corr{the coincidence rate scales as \(1/L^2\)}. \sout{This is taken into account only when the pair rate is directly measured in coincidence, and far from detector saturation.} Details on this method can be found in the supplementary materials.

\corr{With the EPP flux calibrated, the optical fiber is replaced} with a 10mm-thick cuvette \corr{containing} Rhodamine-6G in methanol solution with a concentration of 2mM (panel e). \sout{We find that this concentration yields the strongest TPA-induced fluorescence signal in the present collection geometry.} Fluorescence is collected on a large-area photo multiplier tube (\textit{PMT - Hamamatsu H7421-50}) \corr{oriented at a 90-degree angle from the incident light}. Any scattered IR light is blocked using several shortpass filters.

\section*{Scaling behavior}

\begin{figure}
    \centering
    \includegraphics[width=0.44\textwidth]{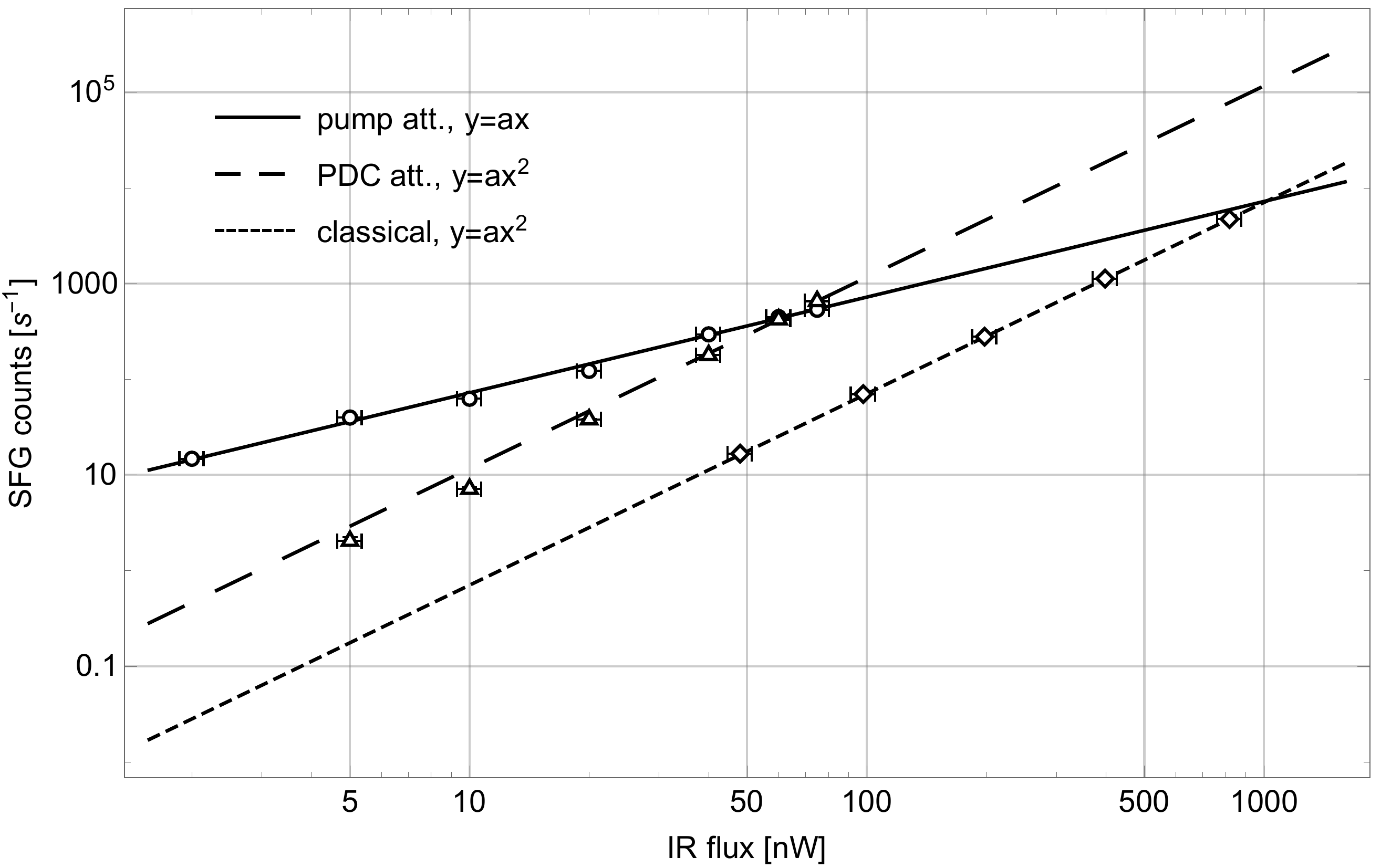}
    \caption{SFG counts at 532\,nm as a function of IR power for three cases: EPP, where power is varied in the parametric down-conversion pump beam (circles), EPP, where power is varied after pair generation (triangles) and a coherent state from a diode laser (diamonds). We show functions corresponding to linear (solid line) and quadratic scaling (dashed and dotted), respectively. Each data point was averaged over 180\,s and had its dark count rate subtracted. \corr{Vertical and horizontal error bars represent shot noise and measurement accuracy of the optical power meter, respectively.}.}
    \label{fig:scaling}
\end{figure}

The first aspect of this investigation is the scaling of efficiency \corr{with input flux} for a \corr{well understood} two-photon process. For this we employ an SFG process in a crystal that is identical with the PDC crystal. In this case with matched nonlinear processes, the SFG is not subject to any spectral enhancement when comparing EPP to CW classical light, because all present wavelength and wave vector combinations are guaranteed to be phasematched. This is the case, since we use broadband avalanche \tsl{photodiodes} and only broadband filtering for the purpose of background suppression. This allows us to probe \corr{directly} the enhancement through photon number correlations in an \corr{independent} manner. For this experiment (panel c in Figure \ref{fig:setup}) we compare EPP and classical CW light. The resulting SFG output count rate as a function of IR power, \corr{representing either the laser light or EPP,} is plotted on a log-log scale in Figure \ref{fig:scaling}. For comparison, we show lines with a slope of 1 and 2, respectively, corresponding to linear and quadratic scaling. \sout{For the case of EPP, we use two different mechanisms to vary IR power: If the EPP is varied directly by attenuation of the IR light between the two nonlinear crystals, the scaling is quadratic. If the IR power is varied by attenuating the SPDC pump power, the SFG output scales linear with IR power. We compare this to classical CW light, where the scaling is again quadratic.}\corra{(Litarally explained in the next paragraph)}

First, varying EPP power directly by attenuation \corr{after the SPDC crystal} verifies that we are in fact probing a two-photon process. Any single-photon process such as one-photon absorption or scattering\sout{, or SPDC pump light bleeding through the filters} would scale linearly. Thus, linear scaling with IR power alone is not a signature of ETPA. However, if instead the SPDC pump power is reduced, the number of produced EPP is reduced while retaining photon number correlations, compared to probabilistically attenuating the photons individually, and scaling with flux is linear \cite{Dayan2005}. \sout{In conclusion, the appropriate way to verify that the process is a two-photon process, and that it is driven by a beam of photon pairs, is to observe the above scaling behavior, ideally in a log-log plot. In the case of competing one- and two-photon processes, or strongly overlapping pairs at high flux, these slopes won't fit exactly. This would be challenging to ascertain in plots with linear scales. For classical light, scaling will always be quadratic. This can also be leveraged as evidence of a two-photon process.}

\sout{For the purpose of enhancing the efficiency of any two-photon process, this result can be interpreted in terms of enhancement at low flux.} As long as optical power is \corr{far} below a certain threshold \sout{- in this case 1000 nW -} SFG \corr{driven by} EPP can provide orders of magnitude higher efficiency than classical CW light. However, the amount of generated SFG in this enhanced regime depends on the interaction strength of the \corr{particular} process involved (which will most likely be weak compared to extraordinarily-polarized type-0 SFG in Lithium Niobate). For a weaker interaction, the generated SFG flux may very well be \sout{at a level of detection signal-to-noise ratio} insufficient for reliable detection.

Showing quadratic scaling \corr{with flux} is essential for probing ETPA, since the signature of TPA in its dependence on pair rate is easily mimicked by small parasitic single-photon processes. Given that we have obtained pure quadratic scaling with pair attenuation, and pure linear scaling with pump attenuation, we can be confident that we are addressing two-photon processes with a beam that contains pairs. This is in line with the past observations from \cite{Georgiades1995,Dayan2005}. \sout{as well as what has recently been described as a best practice in \cite{parzuchowski2020}. In addition, our SFG experiment allows us to minimize second-order dispersion through the use of a prism compressor, which we use to maximize the SFG signal. Details are given in the supplementary materials. This is the second factor that needs to be excluded in order to obtain a definitive bound quantum-enhanced TPA in the absence of a signal.}

\section*{TPA of Entangled Photon Pairs}

Given we ascertained correct \corr{flux} scaling behavior and dispersion compensation, we move on to estimating the magnitude of the enhancement of TPA \corr{by EPP}. To obtain an upper bound analogous to that in \cite{parzuchowski2020}, we need two numbers: The supplied number of photon pairs in the interaction volume, and the minimum detectable fluorescence count rate \corr{without setup}.

To obtain the \corr{usable EPP} rate, we place \corr{the tip of }a single-mode optical fiber (\sout{\textit{Corning HI780C},} core diameter 4\,\(\mu\)m) in place of the Rhodamine sample (compare Figure \ref{fig:setup}d). We focus with a 13\,mm-aspheric lens in order to achieve tight confinement, and align the fiber to the lens. We use a 50:50 fiber beam splitter to probabilistically split pairs, and detect photon pairs in coincidence at two superconducting nanowire single-photon detectors. Their spectral response is flat at 80\,\% within a few percent around 1064nm. After correcting for known sources of loss, namely detector efficiency and attenuation necessary to prevent detector saturation, \corr{we find the }lower bound on the maximum usable pair rate in the fiber is \(2.0(\pm0.2)\times10^9s^{-1}\), measured at a pump power of 1W. \sout{At higher pump power, pair rate drops as a result of what we assume to be photorefractive effects in the PDC crystal.} A complete description of the measurement can be found in supplementary material.

\begin{figure}
    \centering
    \includegraphics[width=0.49\textwidth]{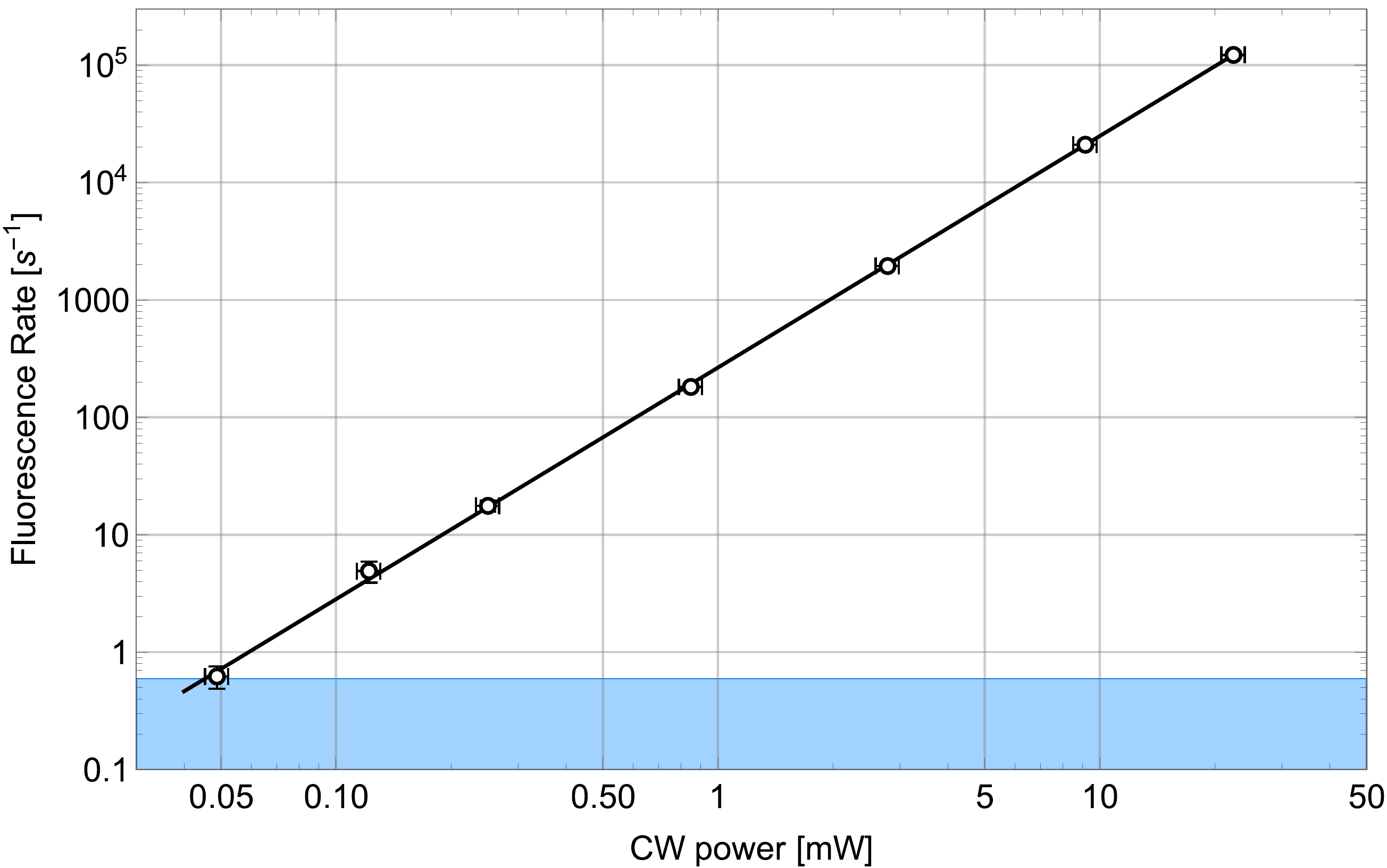}
    \caption{Fluorescence count rates from TPA in Rhodamine-6G. Counts rates were averaged over 5 seconds, and for 1800s for the lowest point. Horizontal error bars reflect measurement accuracy of the optical power meter, vertical error bars represent shot noise. Fit exponent is 1.9725. \corr{The blue shaded area corresponds to fluorescence flux levels below detection threshold.}}
    \label{fig:classicaltpa}
\end{figure}

Next, we \corr{replace} the fiber for a cuvette containing a 2mM solution of Rhodamine-6G. We found this concentration to yield maximum \corr{detected} fluorescence \tsl{from TPA.} We excite a two-photon transition at 1064\,nm with classical light from \corr{the CW IR laser}, and measure the \corr{TPA-induced} fluorescence count rate. The results are shown in Figure \ref{fig:classicaltpa}. At the lowest \corr{usable} power and after subtracting dark counts, we measure a \sout{\corr{TPA-induced} fluorescence} rate of \(0.7(\pm0.1) s^{-1}\). For this data point, we used a chopper wheel to periodically block the incoming light for dark count acquisition with a 50:50 duty cycle, and acquired counts for 3600\,s (1800\,s effective acquisition time \corr{when} the beam was unblocked)\sout{ and subtracted dark counts}.

From these measurements we \sout{can} extract the collection efficiency \(\eta_{\mathrm{col}}\) by fitting the TPA fluorescence rates with a quadratic function \(TPA=a\cdot F^2\) of the flux \(F = P\cdot \lambda/hc\) where 
\(P\) is optical IR power, \(\lambda\) is the laser wavelength, \(h\) is Planck's constant and \(c\) is the vacuum speed of light. From the fit we determine the collection efficiency:
\begin{equation}
    TPA = \frac{F^2\cdot C\cdot\sigma_2}{\pi} \cdot\eta_{\mathrm{col}}\cdot\eta_{\mathrm{det}}\cdot\gamma \cdot \int \frac{1}{\operatorname{w}(z)^2} dz
\end{equation}
where 
\(C\) denotes concentration, \(\sigma_2\) is the \corr{known} TPA cross section \(9.4(\pm1.5)\)GM \cite{Makarov2008}, \(\gamma=0.8\) is the \corr{known} fluorescence yield, \(\eta_{det}=10\%\) is the \corr{known} detection efficiency, and \tsl{\(\textrm{w}(z)\)} is the beam waist along the optical axis. From the fit we obtain a collection efficiency of \(\eta_{col}=1.9(\pm0.2)\%)\), \corr{which is in reasonable agreement with independent estimates for our setup.} This number enables us to estimate the expected fluorescence flux at the maximum EPP pair rate in the fiber. To account for the \corr{expected enhancement of photon number correlations in EPP \corr{relative to} classical CW light (where photons are uncorrelated)}, we introduce a \sout{photon-number-correlation} quantum enhancement factor \(QEF = 2\cdot \sigma_{EPP}/(\sqrt{\pi}\cdot F)=16,000\pm1,000\), where \(\sigma_{EPP}\) denotes the \corr{measured} spectral bandwidth of the EPP (see supplementary material for details):
\begin{equation}
\begin{split}
    TPA_{\mathrm{EPP}} &= QEF\cdot\frac{F^2\cdot C\cdot\sigma_2}{\pi} \cdot\eta_{\mathrm{col}}\cdot\eta_{\mathrm{det}}\cdot\gamma \cdot \int \frac{1}{\operatorname{w}(z)^2}\\
    &= 2.2(\pm0.3)\times10^{-6} s^{-1}
\end{split}
\end{equation}
From the \corr{1800\,s}-long dark count measurement for the observable data point at the lowest power in Figure \ref{fig:classicaltpa} we determine the detection threshold \corr{of our setup} (within a 3-\(\sigma\)-limit) as \(0.7(\pm0.1)s^{-1}\). \corr{The ratio between detection threshold and expected fluorescence rate bounds any additional quantum enhancement (\(E\)) to the value:}
\begin{equation}
    E<\frac{0.7s^{-1}}{2.2\times10^{-6}s^{-1}}=3.2(\pm0.7)\times10^{5}
\end{equation}
Note that our model (see supplementary material for details) accounts for the positive effect of photon number correlations. This additional quantum enhancement factor absorbs any additional enhancement, whether it stems from time-frequency entanglement or otherwise. This measured bound is many orders of magnitude below values reported in the literature, which range from \(10^{10}\) to \(10^{31}\). While this bound would depend on EPP flux, the flux in our experiment is comparable to values used by previous experiments. Any further decrease in flux, which would increase the theoretical enhancement, would only further reduce a signal already below \corr{practical} detection thresholds.

\section*{Discussion}

From the measurements presented we conclude that any enhancement of two-photon absorption by time-frequency entangled photon pairs is many orders of magnitude lower than previously reported. This supports the theoretical predictions in \cite{raymer2020} \corr{that quantum enhancement is equal to \(B/F\), the ratio of EPP bandwidth B to flux F. For our experiment, carried out with CW sources where the enhancement should be maximal, this prediction yields \(B/F=16,000\), falling far short of enabling a successful detection of ETPA using state-of-the-art techniques. It also} corroborates the independent results obtained \corr{under different, less controlled experimental conditions} by Parzuchowski et al. \cite{parzuchowski2020}. While our experimental design can exclude some adverse factors that might in principle diminish TPA, namely lack of spatial overlap, dispersion, and dominating single-photon effects such as scattering, \corr{it should be recognized} that these results are limited to the regime of isolated\corr{, non-overlapping} photon pairs, and that these findings apply only to molecular dyes with broadband absorption spectra. A \sout{brief} discussion on other possible prohibitive effects in observing fluorescence is included in the supplementary materials. Using nonlinear crystal with higher power tolerance such as LBO or KTP would allow us to reach higher pair rates, but would also prompt us to leave the regime of isolated pairs, \corr{which violates the conditions required for creating large a quantum enhancement}. Therefore, we believe our results to be definitive \corr{and represent a best case scenario within their regime of interest}. \corr{This suggests that any observable enhancement beyond what is predicted in \cite{raymer2020} may stem from other effects such as squeezing, novel features in the molecule's electronic structure, or spatial correlations.}

\sout{In the advent of novel molecular systems with high TPA cross sections and fluorescence yield \cite{Bong2009,Dini2008,Tonu2020} it may be possible to better inform more clearly on the actual value of the entanglement enhancement. In the presence of a detectable fluorescence signal a study as a function of the amount of entanglement would become feasible. SFG as a toy model for TPA may be an alternative test bed for such an experiment.}

\section*{Acknowledgements} 

This work was supported by the National Science Foundation RAISE-TAQS
Program (PHY-1839216 to M.G.R., A.H.M., and B.S. as co-PIs).

\bibliography{e-tpa}

\end{document}

%\title{Supplementary Material - Experimental feasibility of molecular two-photon absorption in the isolated photon pair regime}
\title{Supplementary Material}

%\author{Tiemo Landes}
%\email{oqt@uoregon.edu}
%\affiliation{Department of Physics and Oregon Center for Optical, Molecular, and Quantum Science, University of Oregon, Eugene, Oregon 97403, USA}
%\affiliation{Department of Chemistry and Biochemistry, Center for Optical, Molecular and Quantum Science,University of Oregon, Eugene, Oregon 97403, USA}
%\author{Sofiane Merkouche}
%\affiliation{Department of Physics and Oregon Center for Optical, Molecular, and Quantum Science, University of Oregon, Eugene, Oregon 97403, USA}
%\author{Markus Allgaier}
%\affiliation{Department of Physics and Oregon Center for Optical, Molecular, and Quantum Science, University of Oregon, Eugene, Oregon 97403, USA}
%\author{Brian J. Smith}
%\affiliation{Department of Physics and Oregon Center for Optical, Molecular, and Quantum Science, University of Oregon, Eugene, Oregon 97403, USA}
%\author{Andrew H. Marcus}
%\affiliation{Department of Chemistry and Biochemistry, Center for Optical, Molecular and Quantum Science,University of Oregon, Eugene, Oregon 97403, USA}
%\author{Michael G. Raymer}
%\affiliation{Department of Physics and Oregon Center for Optical, Molecular, and Quantum Science, University of Oregon, Eugene, Oregon 97403, USA}

%\date{\today}% It is always \today, today,
             %  but any date may be explicitly specified

%\keywords{Suggested keywords}%Use showkeys class option if keyword
                              %display desired
\maketitle

\section*{Source Parameters and Characterization}

The SPDC source uses a periodically poled, 10\,mm-long Magnesium Oxide-doped Lithium Niobate bulk crystal (\textit{Covesion}) with a poling period of \tslt{6.90}\,\(\mu\)m and a phasematching temperature of 340\,K.
The type-0 process is pumped with CW light from a diode-pumped solid state laser at 532\,nm (\textit{Coherent Verdi V-5}). The forward-propagating (colinear) part of the SPDC mode is collimated with an achromatic lens with a focal length of 100\,.mm. Spatial beam properties, collimation and co-propagation with the 1064\,nm laser beam are verified using a CCD camera at various distances from the collection lens.

The emitted SPDC, centered around 1064\,nm, was sent through \tslt{1000}\,m of optical fiber (\textit{Nufern 780HP}), facilitating a fiber-assisted time-of-flight spectrometer analogous to \cite{avenhaus2009}.

\begin{figure}
    \centering
    \includegraphics[width=0.49\textwidth]{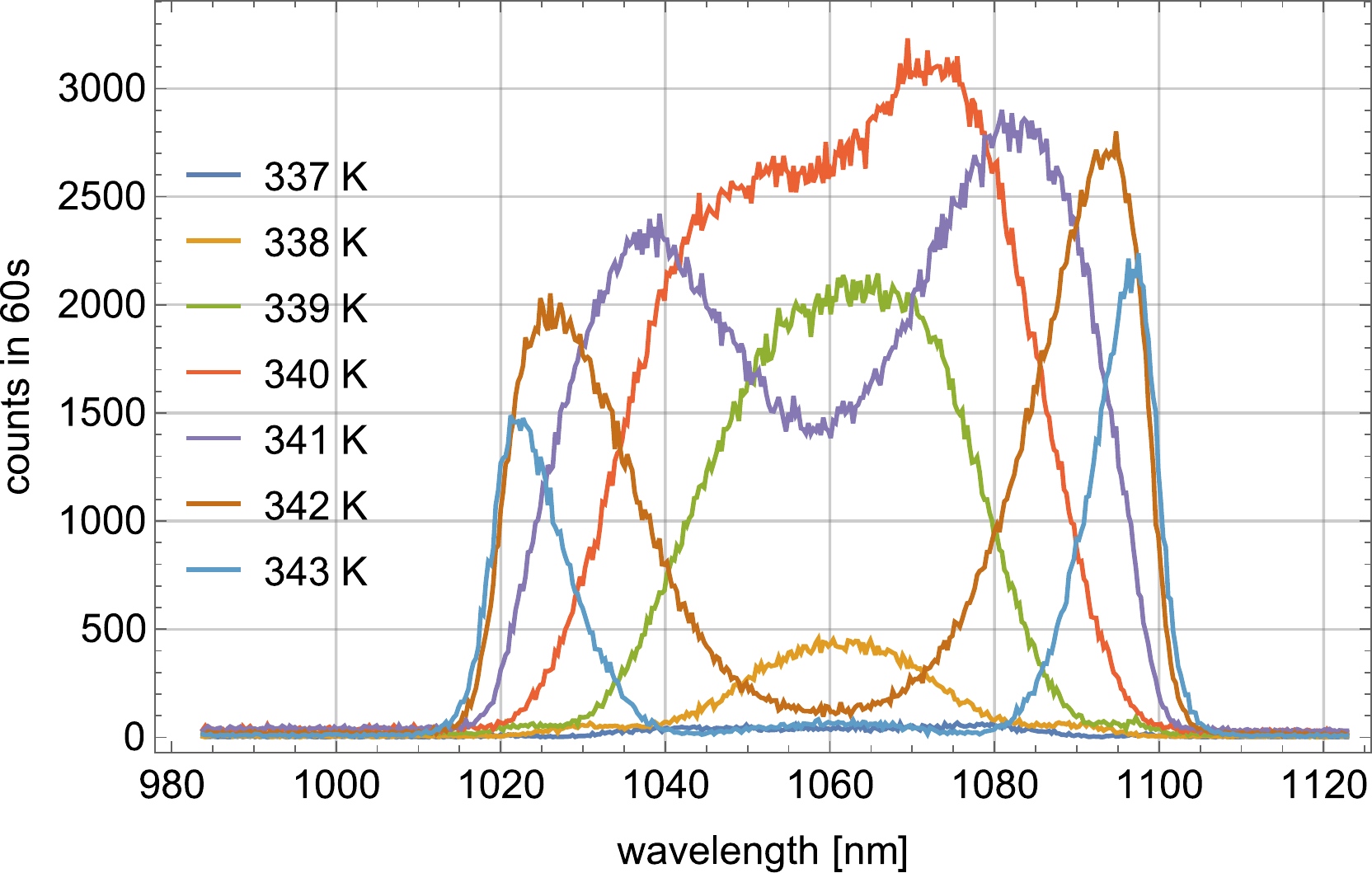}
    \caption{Spectrum of the entangled photon pairs at various crystal temperatures. Spectra were measured on a 500m-long dispersive fiber time-of-flight spectrometer.}
    \label{fig:sup:sourcespec}
\end{figure}

Figure \ref{fig:sup:sourcespec} shows measured EPP spectrum. This includes both photons and thus presents the sum of signal and idler marginal spectra. The FWHM-bandwidth extracted from a Gaussian fit is 40\,nm. Since the shape of CW-pumped type-0 PDC is well known, we can estimate the number of modes \(K\) from the pump beam line width. The pump laser (\textit{Coherent Verdi V-5}) line width is 5\,MHz (RMS) or 6.8\,MHz (FWHM):

\begin{equation}
    K = \sqrt{2}\cdot 40\ \mathrm{nm}\ /\ 6.8\ \mathrm{MHz} = 2.2\cdot10^6
\end{equation}

For the purpose of estimating a lower bound for the number of interacting modes, we placed a single mode fiber (\textit{Corning HI780C}) \tslt{centered} at the location of the front facet of the cuvette. We use a fiber-beam-splitter (\textit{Thorlabs}) to probabilistically split the pairs and count them in coincidence. The output ports of the beam splitter are connected to superconducting nanowire single photon detectors (SNSPDs, \textit{iD Quantique}) with a quantum efficiency of roughly 80\% at 1064nm. Counting the number of pairs allows us to account for any losses in the setup to that point, such as reflection losses on metal mirrors, which wouldn't be possible by inferring the pair rate from measured power at the PDC wavelength.

\begin{figure}
    \centering
    \includegraphics[width=0.49\textwidth]{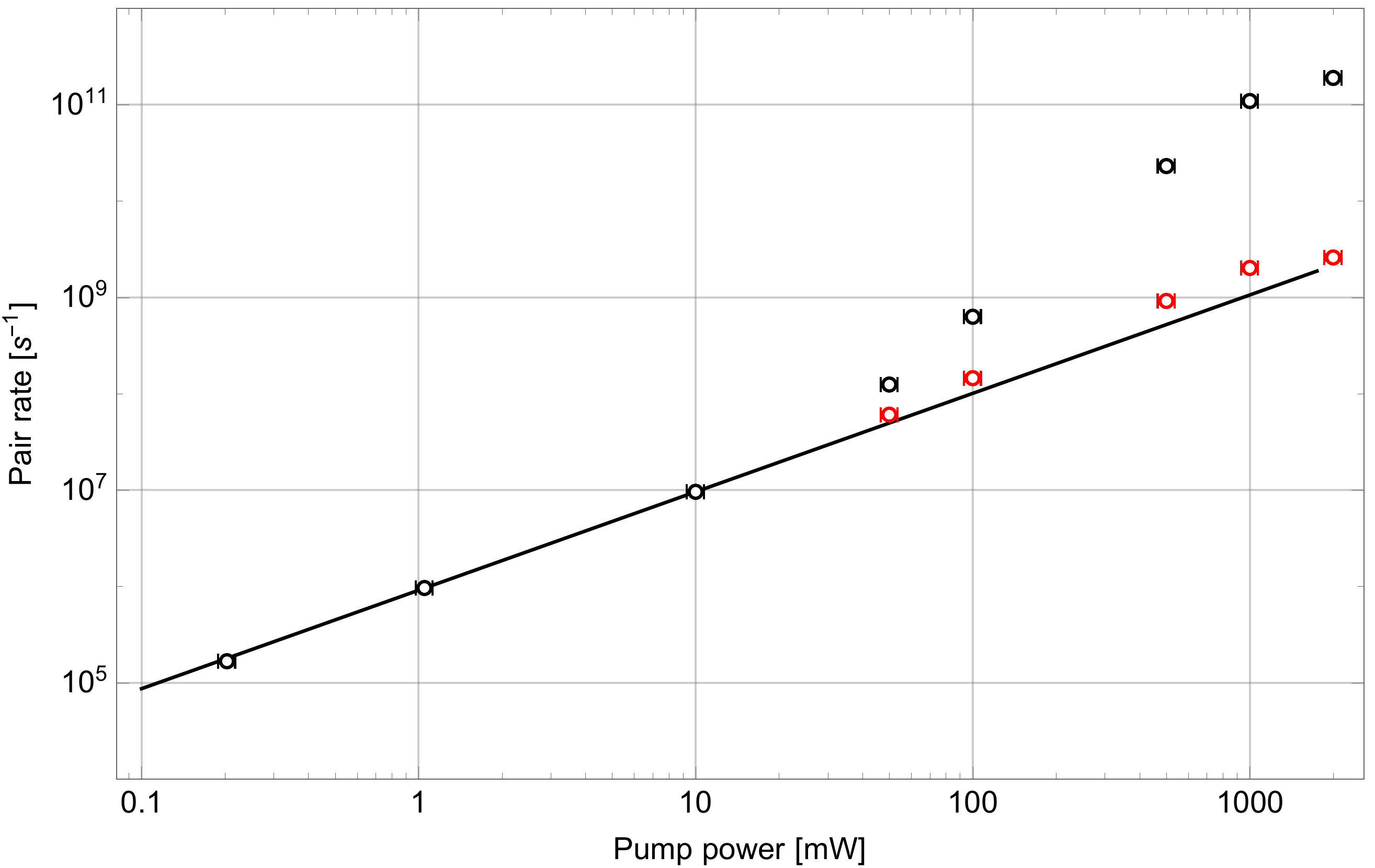}
    \caption{Photon pair coincidence rate as a function of pump power. The measured rate has been adjusted for the attenuation needed to prevent detector saturation. Each data point was averaged over 5 seconds. The solid line represents an exponential fit to the linear regime (first three points), the exponent is 1.02. Points deviating strongly from the linear scaling were excluded and recalculated using Klyshko efficiency in the linear regime, and single count rates (red symbols).}
    \label{fig:sup:pairs}
\end{figure}

We attenuate the EPP to avoid detector saturation, and adjust the coincidence count rates for the attenuation assuming that coincidence counts scale quadratic with attenuation. The adjusted number of counted pairs as a function of pump power is shown in Figure \ref{fig:sup:pairs}. In the regime of isolated photon pairs the scaling is expected to be linear. However, higher photon number components (quadruples, pairs of pairs, etc.) start to dominate our measurement in the presence of high attenuation, which is of the order of \(10^{-5}\) for the highest recorded data point, due to the probabilistic splitting on the 50:50 beam splitter. This causes an unphysically high coincidence count number when corrected for attenuation (larger than the single count rate) and hence the non-linearly scaling points are discarded and recalculated in a different way. We use the linear three data points to calculate the Klyshko efficiency (\(16(\pm2)\%\)), and estimate the high power coincidence rates by multiplying the single count rates with the Klyshko efficiency \cite{klyshko1980}. This data treatment does take into account that the pair generation rate is not linear with power at high power, but avoids overcounting caused by strong attenuation \corr{as well as spectral dependence of attenuation.}

Above 1W of pump power we observe a derivation from linear scaling, which we attribute to photorefraction in the PDC crystal in the shape of diminishing returns in pair production, and we refrained from exceeding that power in our experiments. We measure \(2.0\times10^9\) photon pairs per second at 1\,W pump power. At this rate, the average separation of two pairs is roughly 500\,ps, 3 orders of magnitude larger than their correlation time of 100\,fs.

\section*{Dispersion Compensation}

\begin{figure}
    \centering
    \includegraphics[width=0.25\textwidth]{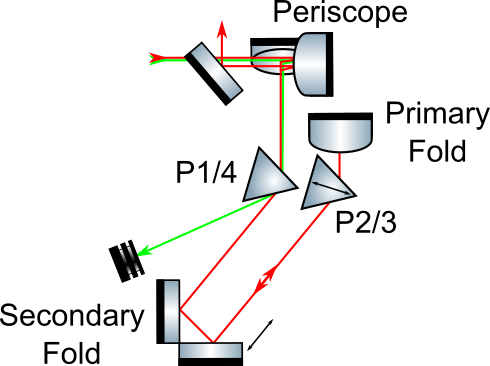}
    \caption{Schematic of the prism compressor. P: Prism.}
    \label{fig:sup:compressor}
\end{figure}

Because of the broadband EPP spectrum, dispersion plays an important role in this experiment. Even small amounts of dispersion in the PDC crystal, lenses and filters, which amount to roughly 4000\,fs\textsuperscript{2}, would cause the photons within each pair to lose overlap drastically. To avoid this, we compensate second order dispersion in an analog fashion to \(\chi^2\)-experiments conducted by Dayan et al. \cite{Dayan2005}. Figure \ref{fig:sup:compressor} shows a more detailed schematic of the employed prism compressor.

In consideration of wavelength, possibility for alignment, and amount of expected dispersion, we chose to use a double-folded prism compressor with SF11-prisms. For zero prism insertion the prism separation would be of the order of 20\,cm. The amount of positive dispersion applied with 10\,mm of prism dispersion would roughly cancel out the negative dispersion achieved by the compressor, allowing for a large degree of variability while guaranteeing stable beam alignment.

To enter the prism compressor and to take advantage of the lack of reflection at the Brewster angle, polarization has to be rotated. Due to the broadband spectrum of the EPP we do this with a periscope.

The prism compressor, consisting of four effective prisms, is folded in two places: Between prism 2 and 3 (main fold), which enables us to use the prism in a double pass, and between prism 1 and 2 (secondary fold), which are effectively also prism 3 and 4, which enables us to change prism separation without realignment. Each fold uses a retro-reflector (roof mirror, or two mirrors oriented at 90 degrees towards each other). The main fold uses a vertical roof mirror. This allows us to keep the incoming and outgoing beam parallel to each other and to the prism axis, but at a different height from the table. This eliminates any asymmetry in the beam mode. The secondary fold uses a horizontal roof mirror on a precision ball-bearing translation rail. The green pump beam does not hit the second prism and is dumped. The beam of the 1064\,nm-diode laser is used for initial alignment. Both prisms sit on turntables and are aligned to minimum deviation using the diode. After alignment, we verified on a CCD camera that the entire PDC beam mode passes the prism compressor and periscope and is unobstructed. We also verified with the camera in several positions that there is no significant beam walk-off due to alignment. An advantage of folding the prism compressor between prism 2\&3 is that and shift caused by changing the insertion of those prisms is inherently cancelled out. This serves as an important tool to verify that changes in signal intensity are caused by dispersion rather than misalignment. We can observe this invariance directly when we measure the number of pairs in a single mode fiber and find that number to be insensitive to prism insertion.

The prism compressor is then finally used to maximize the signal obtained from SFG in the second crystal. SFG is the strongest at minimal dispersion. We adjust this by changing the prism separation using the retro-reflector at the secondary fold. Around the maximum efficiency, we verify that changing prism separation and changing insertion of the second prism (which is inherently stable in terms of alignment) give the same result. We then lock prism separation and insertion in that position before performing all the experiments described in the main text.

\corr{To account for dispersion from the SFG crystal, we have also varied the compressor setting for the ETPA experiment, with no change in its outcome.}

\section*{Expected TPA fluorescence rate from EPP}

To calculate the expected TPA fluorescence flux for excitation with EPP, we use the established theory of classical TPA, which relates fluorescence flux to the excitation flux \(F\):

\begin{equation}
    TPA = \frac{F^2\cdot C\cdot\sigma_2}{\pi} \cdot\eta_{col}\cdot\eta_{det}\cdot\gamma \cdot \int \frac{1}{\operatorname{w}(z)^2}
\end{equation}

where C denotes concentration (in molecules/volume), \(\sigma_2\) is the classical TPA cross section, \(\eta_{col}\) is the collection efficiency, \(\eta_{det}\) is detector efficiency, \(\gamma\) denotes the fluorescence yield, and \(\textrm{w}(z)\) is the beam waist along the optical axis. To adapt this relationship for positive effects of EPP, we employ the definiton of a quantum enhacement factor (QEF) from \cite{raymer2020}. For the relevant case where the two photon transition bandwidth is larger than the bandwidth of the exciting light, the factor reads

\begin{equation}
    QEF = \frac{P_{EPP}}{P_{Coh}}
\end{equation}

where \(P\) denotes the the probability of encountering a pair of photons within a given time window T. For pulsed light, this interval will be the pulse duration. Here, we model CW light as being "chopped" into a series of square pulses of arbitrary duration T, which is defined in a way that forces \(P\ll1\)

For coherent cw light, the  photon probability can be expressed as a function of the flux \(F\):

\begin{equation}
    P_{Coh} = T\cdot F^2
\end{equation}

while for EPP the probability scales linearly with single photon flux (which is twice the pair rate):

\begin{equation}
    P_{EPP} = \frac{2}{\sqrt{\pi}}T\cdot F\cdot  \sigma_{EPP}
\end{equation}

with the spectral EPP bandwdith \(\sigma_{EPP},\ [rad/s]\). The ratio of the two at equal flux describes the ratio of encountering a pair per unit time interval:

\begin{equation}
    QEF = \frac{2\cdot \sigma_{EPP}}{\sqrt{\pi}\cdot F}
\end{equation}

For the comparison that allows us to correct the expected fluorescence flux for the positive effect of the EPP photon number correlation. For a FWHM bandwidth of 40\,nm and a flux of \(2\cdot2.0\times10^{9}\) the enhancement factor is \(16,000\pm1,000\). For a more intuitive picture, consider that the EPP spectral bandwidth is the inverse correlation time \(\tau_c=1/\sigma_{EPP}\). The enhancement factor then reads

\begin{equation}
    QEF = \frac{2}{\sqrt{\pi}\cdot F\cdot \tau_c}
\end{equation}

Hence, the enhancement is the inverse of the probability to encounter a pair of uncorrelated (classical) photons within the correlation time of the EPP. The lower the photon rate, the smaller this probability is, and the larger the enhancement.

\section*{Discussion of adverse effects}

Performing an experiment in the absence of a detectable signal is prone to many pitfalls. Here's we attempt to compile a comprehensive list of such pitfalls, along with our attempts to anticipate them. This list is in parts compiled from Ref. \cite{raymer2020} and \cite{parzuchowski2020} and amended with our own analysis of PDC experiments.

\begin{itemize}
    \item Insufficient spatial overlap - If focusing into the molecular sample is insufficiently tight, or the waist is not inside the sample volume, there is a chance that the two photons within a pair have poor spatial overlap. We address this by using a single-mode fiber as a test volume for our actual photon flux. The pair rate is directly measured in coincidence.
    \item Dispersion - Dispersion would cause the anti-correlated photons in a pair to arrive at different times, and lose temporal overlap, preventing TPA. We estimate the amount of dispersion from the specifications of all employed optics. This allows us to design an appropriate prism compressor that compensates for second order dispersion. We use the SFG experiment to verify successful dispersion compensation. The SMF experiment allows to exclude any adverse effects from alignment caused by adjusting the amount of compression.
    \item Competing single-photon processes. Scattering, pump light bleeding through optical filters, single-photon absorption in the dye, solvent, optics or cuvette, just to name a few, all share the same linear scaling behavior with loss. A linear signature alone is therefore not sufficient to verify that TPA is the dominant process. It is therefore recommended to verify that TPA and fluorescence rates scale quadratically with pair attenuation, as demonstrated in our SFG experiment. A mixture of single and two-photon processes would decrease the detection signal-to-noise ratio, and at least for classical TPA the proper scaling should be verified.
    \item Linear loss - The effect of linear loss on photon pairs is that either photon can be lost probabilistically. Because of this, one cannot infer the average number of pairs from the average number of photons, or optical power. It is therefore necessary to directly measure the number of pairs in coincidence. This allows to ignore any sources of loss between the PDC source and the molecular sample, since the number of photon pairs is known at the sample location.
    \item Detector saturation - Any and all of the signals one may wish to observe here rely on a linear detector response. As the average duration between photon pairs approaches the dead time of the detector, photons arriving in fast succession are undercounted, and detector response to power becomes sub-linear. Dead time of common photon counting detectors are of the order of 50ns for avalanche photo diodes, 10ns for photo multiplier tubes, and as low asfew nanoseconds for superconding nanowire detectors. Therefore, saturation effects usually start to appear at counts in excess of \(10^7\). We attenuate the PDC beam in order to avoid saturation, and then adjust the measured coincidence rate for the amount of attenuation. \corr{At the onset of saturation, correction of detector response may be possible with manufacturer-supplied calibration factors, but only for CW light. For pulsed light with average photon numbers per pulse larger than one, detector dead time correction cannot yield reliable results.}
    \item Insufficient collection efficiency - The most simple explanation for the absence of a fluorescence signal is insufficient collection efficiency, but is easily characterized with classical TPA.
    \item Asymmetric spectral detector response - Silicon-based detectors do not just suffer from low quantum efficiency near the optical band gap of 1100nm, but it is also high asymmetric. This biases detection around 1064nm towards the short-wavelength half of the PDC spectrum. Because of this, optimizing solely on counts can result in poor coupling at the center of the spectrum, so  using raw single-photon count rates to align fiber coupling is not an option. There are three remedies: Use a narrowband spectral filter to align on the center of the spectrum, align directly on coincidences using a fiber beam splitter and two detectors, or use a different sort of detector with a more uniform response. The latter is the case when using nanowire detectors: In the range of 1000-1100nm, their quantum efficiency only varies by few perfect.
    \item \corr{Insufficient EPP generation efficiency - SPDC sources rarely supply all of the flux in the form of pairs. This is typically expressed in the form of Klyshko efficiency \cite{klyshko1980}, which is the ratio of coincidence count rate to single-photon count rate. This quantifies how many usable pairs are contained in a beam emitted from an SPDC source and collected for detection (and experimentation). When the value for the Klyshko efficiency is corrected for detection efficiency, usually referred to as the heralding efficiency, is is typically of the order of 10...80\,\% \cite{harder2013,bock2016}, depending on detection efficiency. The fact that only a fraction of photons in an SPDC beam are actually EPP is another reason to optimize a setup on coincidence counts and measure the number of pairs rather than assume them from power or single counts.}
    \item Reabsorption - The emission and absorption spectra of dyes such as Rhodamine overlap, causing some of the emitted fluorescence photons to be reabsorbed. Higher dye concentration would increase the number of TPA events, but also increase reabsorption of fluorescence photons. \tsl{For our collection apparatus}, using the fluorescence signal from classical TPA, we established that within the range of 0.2mM to 20mM, a concentration of 2mM maximized detection for our collection geometry \tsl{ without appreciable reabsorption}.
\end{itemize}

\bibliography{e-tpa}